\documentclass[10pt,epsf]{article}

\usepackage{graphicx}
\usepackage{indentfirst}

\newcommand{\be}{\begin{equation}}

\newcommand{\ee}{\end{equation}}

\newcommand{\ba}{\begin{array}}

\newcommand{\ea}{\end{array}}

\graphicspath{{converted_graphics/}}
\begin{document}
\begin{titlepage}
\vspace{.5in}
\begin{flushright}
CQUeST-2008-0208
\end{flushright}
\vspace{0.5cm}

\begin{center}
{\Large\bf Geodesic Properties and Orbits in 5-dimensional Hypercylindrical Spacetime}\\
\vspace{.4in}

\end{center}
\begin{center}

  {$\rm{Bogeun \,\, Gwak ,}^{\dag}$}\footnote{\it
  email:rasenis@sogang.ac.kr}\,\,
  {$\rm{Bum-Hoon \,\, Lee ,}^{\dag\P}$}\footnote{\it
  email:bhl@sogang.ac.kr}\,\, and
  {$\rm{Wonwoo \,\, Lee}^{\S}$}\footnote{\it email:warrior@sogang.ac.kr}\,\,

  {\small \dag \it Department of Physics, Sogang University, Seoul 121-742,
  Korea}\\
  {\small \P \it Center for Quantum Spacetime, Sogang University, Seoul 121-742,
  Korea}\\
  {\small \S \it Research Institute for Basic Science, Sogang University, Seoul 121-742,
  Korea}\\

\vspace{.5in}
\end{center}
\begin{center}
{\large\bf Abstract}
\end{center}
\begin{center}
\begin{minipage}{4.75in}

{\small We investigate the geodesic motions of a massive particle
and light ray in the hyperplane orthogonal to the symmetry axis in
the 5-dimensional hypercylindrical spacetime. The class of
the solutions depends on one constant $a$ which is the ratio of
string mass density and tension. There exist unstable orbits in
null geodesic only in some range of $a$. The innermost stable
circular orbits in timelike geodesic also exist only in a certain
range of the parameter $a$. The capture cross section
and the deflection angle of light ray are also computed.}\\
PACS numbers : 04.50.+h, 04.70.-s, 04.90.+e

\end{minipage}
\end{center}
\end{titlepage}

\newpage
\section{Introduction \label{sec1}}

In the study of the hierarchy problem, the extra dimension has
done important roles \cite{nima}. The extra
dimension does not have to be compactified 5th dimension. Our
universe can be viewed as a 4-dimensional brane world embedded in the
5-dimensional spacetime. From the analysis of brane world with
large extra dimension, interesting 5-dimensional black object
solutions have been presented \cite{hawk01}. The simplest
model may be the Schwarzschild black string, whose instability was
also examined \cite{gl01}.

General class of solutions including
the Schwarzschild black string were discovered by Kramer
\cite{kra01}, Gross and Perry \cite{gp01}, Davidson and Owen
\cite{do01}, and Lee \cite{chul}. This class of the 5-dimensional
solutions depends on an arbitrary constant $a$. This solution has a naked singularity, and we need to consider the physical possibility of the metric. Virbhadra and Ellis \cite{vire} classified naked singularities into weakly and strongly naked ones. There exist parameter range in which singularity is weakly and strongly naked one in our analysis. According to their classification, the singularity of the metric studied in the present work and in the Refs.\ \cite{chul, ckkl, ckkl1, hjko} corresponds to a weakly naked one. Virbhadra and Ellis studied the possibility for observing the naked singularity based on the gravitational lensing effect, pointing out that the cosmic censorship hypothesis of Penrose has not been completely proven. It was also shown by K. S. Virbhadra and C. R. Keeton in Ref.\ \cite{virk} that this observational method may play the role of more efficient cosmic telescopes, if singularities exist in nature. Virbhadra and Ellis also pointed out that the observational properties of the weakly naked singularity as in the metric in our paper will be shown to be similar to those of the Schwarzschild black hole. With all these and in relation to the observation of the singularity in the future, it is worthwhile to investigate the properties of our metric, as has been done by authors in Refs.\ \cite{chul, ckkl, ckkl1, hjko}.

In Ref.\
\cite{chul}, this constant $a$ is interpreted as the ratio of
string mass density and tension. This solution corresponds to the
Schwarzschild black string for $a=\frac{1}{2}$. In addition, the
solution becomes the Kaluza-Klein bubble \cite{eho01} for $a=2$.
Extension of the string to the p-brane solutions with mass and
tension were constructed in Ref.\ \cite{lkk}. In Ref.\ \cite{ckkl},
the authors studied some geometrical properties such as the location of
curvature singularity, proper length of 5 dimension, and casual
structure based on the radial null geodesics. The hypercylindrical solution is asymptotically
flat in both $\rho\sim\infty$ and $\rho\sim 0\,$ \cite{ckkl1}\,. 

The stationary vacuum cylindrical solution with momentum in
translational direction were investigated \cite{lk02}. In this
solution it was shown that the compactification along
$z$-direction breaks the Lorentz symmetry, which makes the frame
dragging effect into the physical observable \cite{lk01}. It was shown
that there are vacuum stationary black string solutions, other than
the Schwarzschild black string solution in 5-dimensional spacetime \cite{kl01}.
These can lead to a new class of solutions, and its properties were
studied in Ref.\ \cite{hm01,cd01,cle01}.
The formation of five-dimensional solutions \cite{chul} from the
gravitational collapse and its numerical works are presented in
Ref.\ \cite{hjko}. A black string from $D$- and $DF$-strings in a
$D3\bar{D}3$ system are studied in Ref.\ \cite{kkkl}. These
objects may play the role of gravitational lenses with
modifying the properties \cite{resl01, virk}.

Due to the complexity of this solution, much of the properties are
not yet known. In this paper we study the geodesics and the orbits
according to $a$.

The organization of this paper is as follows. In Sec.\ 2, we
briefly review the hypercylindrical solutions. We summarize the
physical meaning of the constant $a$, the condition for event
horizon, the Schwarzschild black string, and the Kaluza-Klein
bubble. In Sec.\ 3, the main results are calculated \cite{gwak}.
We derive the geodesics for null and timelike from Lagrangian, and
get conserved quantities using Killing vectors. The range of $a$
giving rise to unstable circular orbit is obtained by analyzing
the effective potential. The diffraction angle of bending
light is also studied. Timelike geodesics are analyzed by analytic
and numerical methods. In addition, some properties of orbits are
numerically computed. In Sec.\ 4, we summarize the results with
discussion.

\section{Brief Review on the hypercylindrical Solutions  \label{sec2}}

A general class of the hypercylindrical solutions with an
arbitrary constant $a$ is found and explained in Ref.\
\cite{chul}. The tension of the string $\tau$ and its mass density
$\zeta$ are related by
\begin{equation}
\tau=a \zeta\,.
\end{equation}
In Ref.\ \cite{traschen}, the author studied the positive tension due to
pure gravitational contribution. In addition, if one demands that
the strong energy condition be satisfied, there exists the upper
bound, $a \le 2$ \cite{ckkl}. We will consider the general case including
the negative tension.

The metric ansatz for hypercylindrical solutions is written as
\begin{equation}
{ds}^2=-F(\rho ){dt}^2+G(\rho )({d\rho }^2+\rho ^2{d\theta}^2+\rho
^2\sin ^2{\theta d\phi }^2)+H(\rho ){dz}^2\,. \label{02-metric01}
\end{equation}
The components of the metric satisfying the asymptotical flatness are \cite{chul}
\begin{eqnarray}
F&=&\left(1-\frac{K_a}{\rho }\right)^s\left(1+\frac{K_a}{\rho
}\right)^{-s}\,, \,\,\,\,\,\,  G=\left(1-\frac{K_a}{\rho
}\right)^{2-\frac{(1+a) s}{2-a}}\left(1+\frac{K_a}{\rho
}\right)^{2+\frac{(1+a) s}{2-a}}\,, \nonumber \\
H&=&\left(1-\frac{K_a}{\rho }\right)^{\frac{(-1+2 a)
s}{2-a}}\left(1+\frac{K_a}{\rho }\right)^{\frac{(1-2 a) s}{2-a}}\,,
\nonumber
\end{eqnarray}
where $s=\frac{2(2-a)}{\sqrt{3(1-a+a^2)}}$ and
$K_a=\sqrt{\frac{1-a+a^2}{3}}G_5\zeta$\,.

The metric has the transverse spherically symmetric static property. We will consider the range of $\rho$, $K_a\leq\rho<\infty$\,.

In the above solutions, $a$ is an arbitrary constant not depending
on coordinate variables. The causal structure of this spacetime is
described in Ref.\ \cite{ckkl} by analyzing the radial null
geodesics. In that paper, they show that an event horizon exists
only for the case $a=\frac{1}{2}$, which correspond to the
Schwarzschild black string. For other values of $a$, the metric becomes singular at $\rho=K_a$, as pointed out in Ref.\ \cite{lkk}. This singularity corresponds to a weakly naked one in Ref.\ \cite{vire}.

As some special cases, we consider four cases of $a$, which give
different types of solutions. For the constant
$a=0$, the string tension vanishes, and the metric
becomes
\begin{eqnarray}
ds^2&=&-\frac{(1-\frac{K_0}{\rho})^{4/\sqrt{3}}}{(1+\frac{K_0}{\rho})^{4/\sqrt{3}}}{dt}^2
+\left(1-\frac{K_0}{\rho}\right)^{2-2/\sqrt{3}}\left(1+\frac{K_0}{\rho
}\right)^{2+2/\sqrt{3}}({d\rho}^2+\rho ^2{d\theta}^2+\rho^2\sin
^2{\theta d\phi}^2) \nonumber \\
&+&\frac{(1-\frac{K_0}{\rho })^{2/\sqrt{3}}}{(1+\frac{K_0}{\rho
})^{2/\sqrt{3}}}{dz}^2\,,
\end{eqnarray}
where $K_0=\frac{1}{\sqrt{3}}G_5\zeta$\,.

For value of $a=\frac{1}{2}$\,, the metric becomes that of the
Schwarzschild black string given by
\begin{equation}
{ds}^2=-\frac{(1-\frac{K_{1/2}}{\rho})^2}{\left(1+\frac{K_{1/2}}{\rho
}\right)^2}{dt}^2+
\left(1+\frac{K_{1/2}}{\rho}\right)^4({d\rho}^2+\rho^2{d\theta}^2+\rho^2
\sin^2\theta d\phi^2)+{dz}^2\,, \label{02-schbs}
\end{equation}
where $K_{1/2}=\frac{1}{2}G_5\zeta$\,.

For $a=1$\,, the metric is given as follows
\begin{equation}
ds^2=-\frac{(1-\frac{K_1}{\rho})^{2/\sqrt{3}}}{(1+\frac{K_1}{\rho})^{2/\sqrt{3}}}dt^2
+\frac{(1+\frac{K_1}{\rho})^{2+4/\sqrt{3}}}{(1-\frac{K_1}{\rho})^{-2+4/\sqrt{3}}}
(d\rho^2+\rho^2d\theta^2+\rho^2\sin^2d\phi^2)+\frac{(1-\frac{K_1}{\rho})^{2/\sqrt{3}}}
{(1+\frac{K_1}{\rho})^{2/\sqrt{3}}}dz^2\,,
\end{equation}
where $K_1=\frac{1}{\sqrt{3}}G_5\zeta$\,. Note that the metric
functions $g_{tt}$ and $g_{zz}$ are the same.

For $a=2$\,, the solution becomes that of the static Kaluza-Klein
bubble. The metric takes the form
\begin{equation}
ds^2=-dt^2+\left(1+\frac{K_2}{\rho}\right)^{4}
(d\rho^2+\rho^2d\theta^2+\rho^2\sin^2d\phi^2)+
\frac{(1-\frac{K_{2}}{\rho})^2}{\left(1+\frac{K_{2}}{\rho
}\right)^2}dz^2\,,
\end{equation}
where $K_2=\frac{1}{\sqrt{3}}G_5\zeta$\,. This spacetime can be
constructed by the double Wick rotation of the Schwarzschild black
string in Eq.\,(\ref{02-schbs}) \cite{eho01}.

\section{Geodesic Properties and Orbits  \label{sec3}}

One define Lagrangian as
\begin{equation}
\mathcal{L}=\frac{1}{2}g_{\mu\nu}\dot{x}^{\mu}\dot{x}^{\nu}\,,
\label{03-lagrangian}
\end{equation}
where $\cdot$ denotes the differentiation with respect to the affine parameter
$\lambda$. For timelike, null and spacelike case, $\mathcal{L}$ is
equal to -1, 0 and 1, respectively.

The geodesic equation is given by the Euler-Lagrange equation
\begin{equation}
\frac{\partial \mathcal{L}}{\partial
x^{\mu}}-\frac{d}{d\lambda}\frac{\partial \mathcal{L}}{\partial
\dot{x}^{\mu}}=0\,. \label{03-eula}
\end{equation}

The metric is invariant under translations $t \rightarrow t+\Delta t$\,, $\phi
\rightarrow \phi +\Delta \phi$\,, and $z \rightarrow z+ \Delta z$\,.
Accordingly, the conjugate momenta $p_0\equiv -E$, $p_3\equiv \pm
L$, and $p_4\equiv \pm W$ are conserved. Using the Killing vectors
$\xi_0$, $\xi_3$, and $\xi_4$ from the metric
symmetries of $t$, $\phi$, and $z$, respectively, the conserved quantities are expressed as
\begin{eqnarray}
p_0&=& \xi_0\cdot u=g_{\mu\nu}\xi^\mu
u^\nu=-F\frac{dt}{d\lambda}=-E\,, \,\,\,\,\,\, p_3=+\xi_3\cdot
u=g_{\mu\nu}\xi^\mu u^\nu=G\rho^2\frac{d\phi}{d\lambda}=L\,,
\nonumber \\
p_4&=& \xi_4\cdot u=g_{\mu\nu}\xi^\mu \cdot
u^\nu=H\frac{dz}{d\lambda}=W\,. \nonumber
\end{eqnarray}
From Lagrangian, we get the following equations
\begin{equation}
\dot{t}=\frac{E}{g_{00}}=\frac{E}{F}\,, \,\,\,\,
\dot{\phi}=\frac{L}{g_{33}}=\frac{L}{\rho^2G}\,, \,\,\,\,
\dot{z}=\frac{W}{g_{44}}=\frac{W}{H}\,, \label{03-kvpmc01}
\end{equation}
where the conversed quantities $E$, $L$, and $W$ are the energy, angular
momentum, and linear momentum, respectively. In general,
this kind of metric has additional conserved quantities such as
$L^2$. The setting on the equatorial plane,
$\theta=\frac{\pi}{2}\,,$ is related to the remaining two Killing vectors, which lead to the conservation of the direction of angular momentum. Thus, in this setting, conserved quantities are three: $E$, $L$, and $W$, which are integration
constants. In this paper, we can take
$\theta=\frac{\pi}{2}$ without loss of generality, and we investigate the case of no 5-dimensional momentum $W=0$ to see 4-dimensional cases.

To obtain the orbits in this spacetime, we construct the effective
potential. The potential can be obtained from Lagrangian in Eq.\
(\ref{03-lagrangian}). Using Eq.\ (\ref{03-kvpmc01}), the
Lagrangian in Eq.\ (\ref{03-lagrangian}) becomes
\begin{equation}
\mathcal{L}=-\frac{E^2}{F(\rho)}+\dot{\rho}^2G(\rho)
+\frac{L^2}{\rho^2G(\rho)}+\frac{W^2}{H(\rho)}\,.
\end{equation}
This can be rewritten as
\begin{equation}
\frac{1}{2}M_{eff}(\rho)\dot{\rho}^2
+V_{eff}(\rho)=\frac{1}{2}E^2\,,
\end{equation}
where $
M_{eff}=F(\rho)G(\rho)$ and
$V_{eff}=\frac{F}{2}(\frac{L^2}{\rho^2G(\rho)}+\frac{W^2}
{H(\rho)}-\mathcal{L})\,.$ This equation governs the geodesics in
the given spacetime. In the analysis of the geodesic motions, we
can directly read off the behaviors of the motions from the shape
of the effective potential. Stable circular orbits are possible if
the potential has a minimum point. On the other hand, unstable
circular orbits are possible if the potential has a maximum point.
For example, there are stable circular orbits in the case of
timelike geodesics in the Schwarzschild black hole spacetime,
while there are unstable circular orbits in the case of null
geodesics in the spacetime \cite{misner}.

\begin{figure}[h]
\centering
\includegraphics[bb=88 4 376 182,scale=0.6,keepaspectratio]{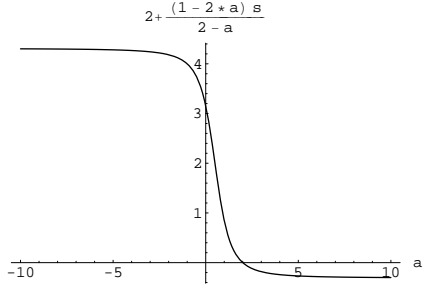}
\caption{The exponent's behavior in $-10<a<10$. It change the sign
at $a=2$.}
\label{fig:massarange1}
\end{figure}

\begin{figure}[h]
\centering
\includegraphics[bb=88 4 376 182,scale=0.5,keepaspectratio]{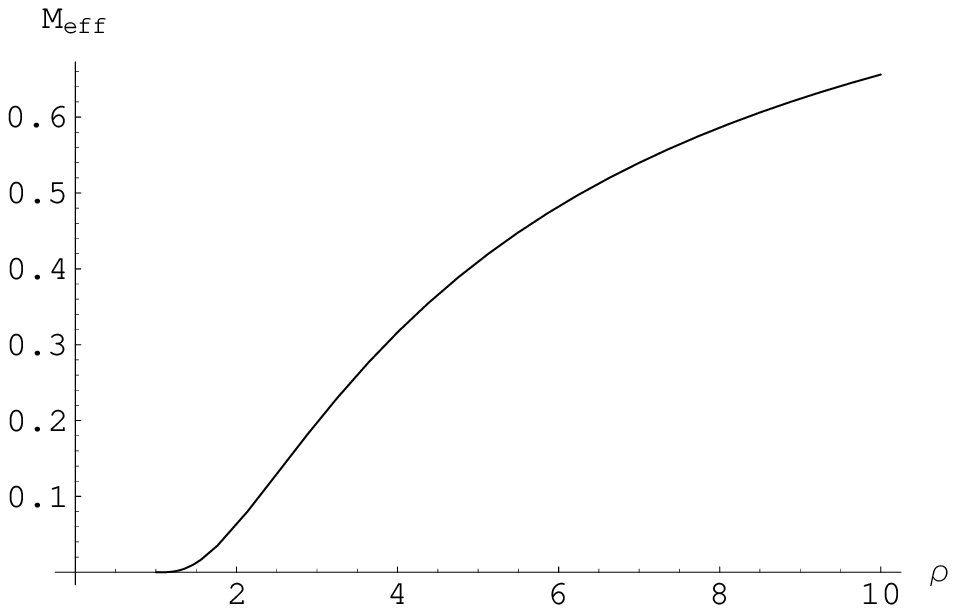}
\includegraphics[bb=88 4 376 182,scale=0.5,keepaspectratio]{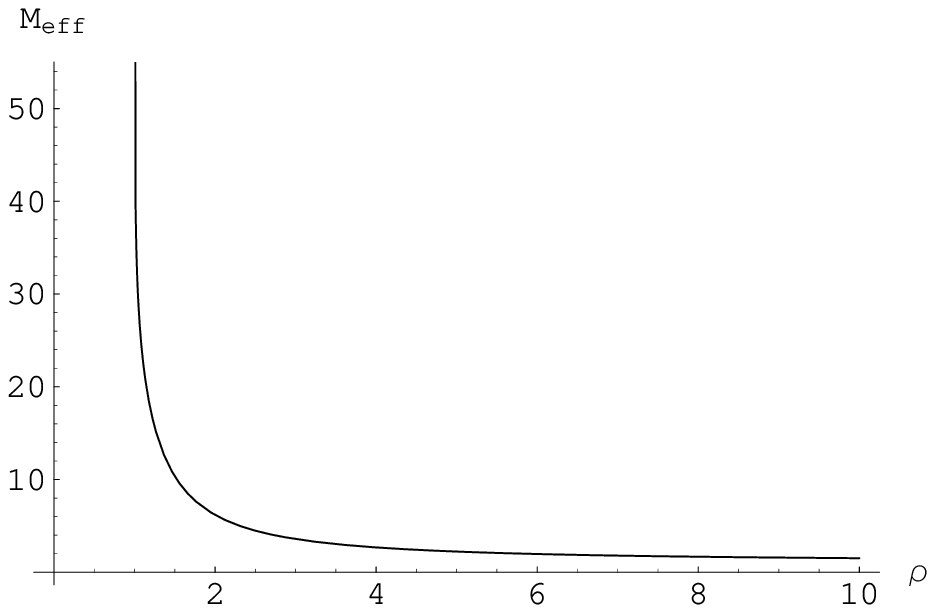}
\caption{Two type of effective mass. The left is for $a<2$. The
right is for $a > 2$.}
\label{fig:masstype1}
\end{figure}

The effective mass is written
\begin{eqnarray}
M_{eff} & = & F(\rho)G(\rho)\nonumber\\
& = & \left(1-\frac{K_a}{\rho}\right)^{2+\frac{1-2a}{2-a}s}\left(1
+\frac{K_a}{\rho}\right)^{2-\frac{1-2a}{2-a}s}\,.\label{mass1}
\end{eqnarray}
Objects which is located in $\rho \gg K_a$ have different mass
called effective mass as the location is approached to $K_a$. It
makes that the object is looked like having mass even if light.
The effective mass is categorized two types by its behavior in the
vicinity of $K_a$.
The effective mass depends on the location $\rho\,$, approaching
asymptotically 1 in any choice of $a$. The behavior of effective mass is divided by
$\left(1-\frac{K_a}{\rho}\right)$'s exponent. The exponent with the base $\left(1-\frac{K_a}{\rho}\right)$
change sign at $a=2$ (Fig.\ \ref{fig:massarange1}). The effective mass
at $\rho=K_a$ becomes zero in $a<2$, and infinite in $a > 2$ (Fig.\ \ref{fig:masstype1}).

\subsection{The Null Geodesics}

The Lagrangian for null geodesics becomes
\begin{equation}
\mathcal{L}=\frac{E^2}{g_{00}}+g_{11}(\dot{\rho})^2+
\frac{L^2}{g_{33}}+\frac{W^2}{g_{44}}=0\,. \label{03-ngc01}
\end{equation}
The geodesic equation of $\rho$ for light ray is given by
\begin{equation}
g_{00}'\left(\frac{E}{g_{00}}\right)^2-g_{11}'(\dot{\rho})^2+g_{33}'\left(\frac{L}{g_{33}}\right)^2
+g_{44}'\left(\frac{W}{g_{44}}\right)^2-2g_{11}\ddot{\rho}=0\,.
\label{03-ngc02}
\end{equation}
From Eq.\ (\ref{03-ngc01})
\begin{equation}
\dot{\rho}=\left(-\frac{1}{g_{11}}\left(\frac{E^2}{g_{00}}+\frac{L^2}{g_{33}}
+\frac{W^2}{g_{44}}\right)\right)^{\frac{1}{2}}\,.
\label{03-rhongc01}
\end{equation}
After inserting this result into the geodesic equation Eq.\,(\ref{03-ngc02}), we get
\begin{equation}
\ddot{\rho}=\frac{1}{2g_{11}}\left(\frac{g_{00}'}{g_{00}}\frac{E^2}{g_{00}}
+\frac{g_{00}'}{g_{11}}\left(\frac{E^2}{g_{00}}+\frac{L^2}{g_{33}}+\frac{W^2}{g_{44}}\right)
+\frac{g_{33}'L^2}{g_{33}^2}+\frac{g_{44}'L^2}{g_{44}^2}\right)\,.
\end{equation}
From Eq.\ (\ref{03-kvpmc01}) and Eq.\ (\ref{03-rhongc01}), we get
the relations between coordinate components as follows
\begin{eqnarray}
\frac{dt}{d\rho}&=&\frac{\frac{E}{g_{00}}}{(-\frac{1}{g_{11}}(\frac{E^2}{g_{00}}
+\frac{L^2}{g_{33}}+\frac{W^2}{g_{44}}))^{\frac{1}{2}}}\,, \,\,
\nonumber \\
\frac{d\phi}{d\rho}&=&\frac{\frac{L}{g_{33}}}{(-\frac{1}{g_{11}}
(\frac{E^2}{g_{00}}+\frac{L^2}{g_{33}}+\frac{W^2}{g_{44}}))^{\frac{1}{2}}}
\,, \,\, \frac{dz}{d\rho}=\frac{\frac{W}{g_{44}}}{(-\frac{1}{g_{11}}
(\frac{E^2}{g_{00}}+\frac{L^2}{g_{33}}+\frac{W^2}{g_{44}}))^{\frac{1}{2}}}
\,. \nonumber
\end{eqnarray}

\begin{figure}[h]
\centering
\includegraphics[bb=88 4 376 182,scale=0.75,keepaspectratio]{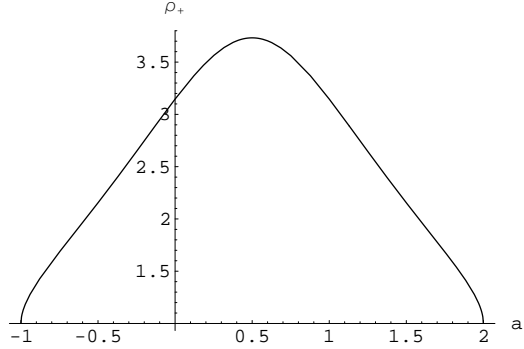}
\caption{The graph of the maximum points of $V_{eff}$. This graph
indicates that the location of the possible unstable circular
orbit is influenced by the constant $a$. The $\rho$ has meaning in
the range, Eq.\ (\ref{03-range01}). In addition, the maximum point
of the graph is at $a=\frac{1}{2}$, which is the case of the
Schwarzschild black string. } \label{fig:extremrho}
\end{figure}

\begin{figure}[h]
\centering
\includegraphics[bb=88 4 376 182,scale=0.6,keepaspectratio]{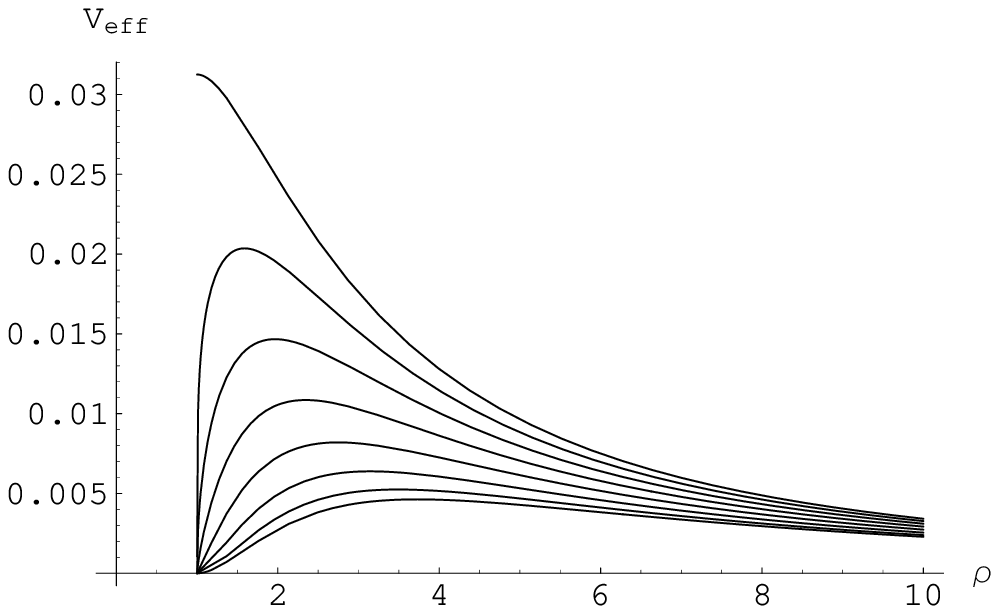}
\includegraphics[bb=88 4 376 182,scale=0.6,keepaspectratio]{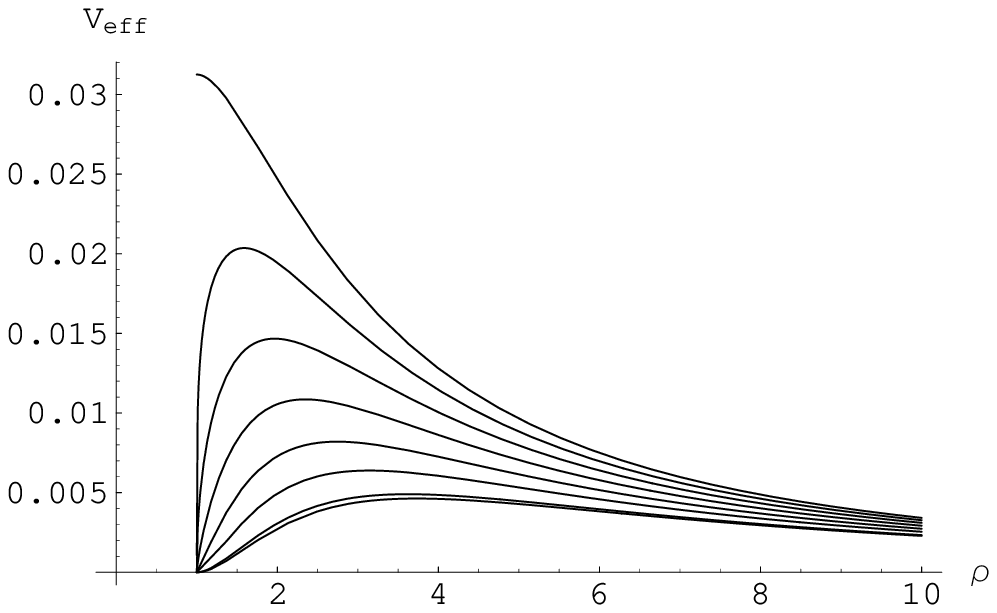}
\caption{The effective potential from $a=-1$ to
$a=0.5$ (left) and from $a=0.5$ to $a=2$ (right). In the left
, the extremum of $V_{eff}$ is located at $\rho_{+}=K_a$. The
extremum value of $V_{eff}$ becomes lower and the location of the
extremum point of $\rho_+$ increases as $a$ increases for $-1\le a\le \frac{1}{2}$. The right graph indicates
that the extremum value of $V_{eff}$ becomes higher and the
location of the extremum point of $\rho_+$ decreases as $a$
increases for $\frac{1}{2} \le a\le 2$.}
\label{fig:allplotm1}
\end{figure}

We consider the orbits of null geodesics satisfying $\dot z
=0$ and $\theta=\frac{\pi}{2}$. The effective potential,
$V_{eff}$, of the 4-dimensional null Lagrangian becomes
\begin{eqnarray}
V_{eff}&=&\frac{1}{2}\frac{F L^2}{\rho^2G(\rho)} \nonumber\\
&=&\frac{L^2}{2\rho^2}\left(1-\frac{K_a}{\rho}\right)^{-2+s+\frac{1+a}{2-a}s}\left(1
+\frac{K_a}{\rho}\right)^{-2-s-\frac{1+a}{2-a}s}\,.
\end{eqnarray}
The effective potentials according to $a$ are
shown in Fig.\ \ref{fig:allplotm1}. The maximum point in the
potential indicate the location of the unstable circular orbit.  The photon spheres corresponds to the location of maximum points in the effective potential for each values of the constant $a$. The surface of photon spheres contain the region of $\rho=K_a$. This type of singularity is called a weakly naked singularity.
Note that $\rho$ can not be smaller than $K_a$.

The angular momentum $L$ appears as scaling factor of
this potential, this is from the fact that the potential expresses
the light motion. The constant $a$ is the only variable
determining the shape of the potential. To get unstable circular
orbits in the potential, we calculate the extrema of the
potential. The location is given by
\begin{equation}
\rho_{\pm}=\frac{(s+\frac{1+a}{2-a}s)\pm
\sqrt{(s+\frac{1+a}{2-a}s)^2-4}}{2}K_a\,,
\end{equation}
where this is the only possible circular orbit.

From the reality of $\rho_{+}$, the range of $a$ is given by
\begin{equation}
-1\le a\le 2\,. \label{03-range01}
\end{equation}
The location of the unstable circular orbit is dependent on
the constant $a$. The behavior of the location can
be read off directly from the Fig.\ \ref{fig:allplotm1}. Note that there are no stable circular photon orbits in this spacetime for
$-1\le a\le 2$ like the case of the Schwarzschild black
hole. There exist parameter range in which singularity is weakly naked one, $-1<a<2$, and strongly naked one, $a\leq-1$ or $a\geq2$, in our analysis.

\begin{figure}[h]
\centering
\includegraphics[bb=88 4 376 182,scale=0.7,keepaspectratio]{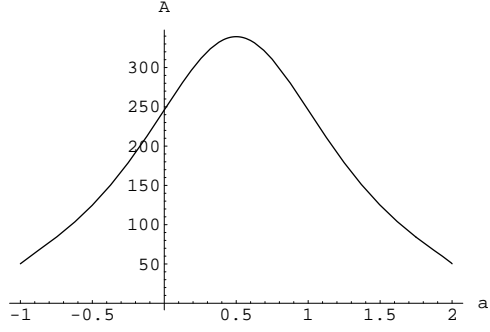}
\caption{The graph of the area in the range of $-1\le a\le2\,$. The point
which the area becomes maximumat $a=\frac{1}{2}\,$.}
\label{fig:area}
\end{figure}

The light ray with impact parameter smaller than specific value is
captured by the strong gravitational field. Using this parameter,
the capture cross section can be calculated. From trajectories of
$\rho$ and $\phi$, 4-dimensional Lagrangian
can be written as
\begin{equation}
\frac{1}{2}\left(\frac{d\rho}{d\phi}\right)^2\frac{F(\rho)}{\rho^4G(\rho)}
+\frac{1}{2}\left(\frac{F(\rho)}{\rho^2G(\rho)}\right)=\frac{1}{2}b^{-2}\,,
\end{equation}
where $b$ is impact parameter. From above equation, the potential impact parameter $B(\rho)$ can
be defined as
\begin{equation}
B(\rho)=\left(\frac{F(\rho)}{\rho^2G(\rho)}\right)^{-\frac{1}{2}}\,.
\label{03-brho01}
\end{equation}
This potential impact parameter is interpreted as the condition
restricting the evolution of null geodesic \cite{misner}
\begin{equation}
b\le B(\rho)\,. \nonumber
\end{equation}
Thus, the minimum value of $B(\rho)$ is the critical impact
parameter, $b_{crit}$. For example, the value of critical $b$ is
$3\sqrt{3}M$ in the Schwarzschild case. Since the shape of $B(\rho)$ is
relative to the inverse $V_{eff}$. The minimum point of
$B(\rho)$ is the same as maximum point of $V(\rho)$. To get the
minimum point, the partial derivative $B(\rho)$ is needed
\begin{equation}
\partial_{\rho}B(\rho)=0 .
\end{equation}
Similar analogy can be done. The same result can be obtained by $B(\rho)$.
The critical value of
impact parameter is $[B(\rho)]_{min}$. Explicitly, $\rho_{+}$ is
redefined to $\rho_{min}$. For $-1\le a \le 2$, the $b_{crit}$ is
written as
\begin{eqnarray}
b_{crit}&=&\rho_{min}\left[\left(1-\frac{K_a}{\rho_{min}}\right)^{-2+s+\frac{1+a}{2-a}s}\left(1
+\frac{K_a}{\rho_{min}}\right)^{-2-s-\frac{1+a}{2-a}s}\right]^{-\frac{1}{2}}, \\
\rho_{min}&=&\frac{1}{2}\left[s+\frac{1+a}{2-a}s+\sqrt{\left(s+\frac{1+a}{2-a}s\right)^2-4}\right]K_a\,,
\nonumber \\
s&=&\frac{2(2-a)}{\sqrt{3(1-a+a^2)}}\,. \nonumber
\end{eqnarray}
For example, the incident light ray whose impact parameter is
smaller than $b_{crit}$ is captured by the gravity of the black hole. The area
exists for $-1\le a\le2\,$.
The area is the capture cross section \cite{raine}, $\pi b_{crit}^2\,$, is given by

\begin{equation}
A=\pi\rho_{min}^2\left[\left(1-\frac{K_a}{\rho_{min}}\right)^{-2+s+\frac{1+a}{2-a}s}
\left(1+\frac{K_a}{\rho_{min}}\right)^{-2-s-\frac{1+a}{2-a}s}\right]^{-1}\,.
\end{equation}
The graph is plotted in Fig.\
\ref{fig:area}.

We now calculate the bending of light in the gravitational field.
To see this bending, we get the trajectory of light rays, using the symmetry of the spacetime.

\begin{figure}[tbp] 
  \centering
  \includegraphics[bb=88 4 1104 726,scale=0.2,keepaspectratio]{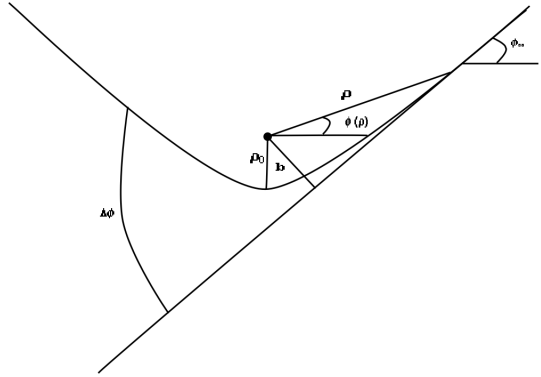}
  \caption{The geometrical figure of the bend of light. The motion
is considered in the equatorial plane, $\theta=\frac{\pi}{2}$.}
  \label{fig:figu1}
\end{figure}

The closest distance $\rho_0$ is the minimum point of the
function, $\rho(\phi)$. The equation
\begin{equation}
\frac{d\rho}{d\phi}=\frac{\rho^2G(\rho)^{\frac{1}{2}}}{L}
\left(\frac{E^2}{F(\rho)}-\frac{L^2}{\rho^2G(\rho)}\right)^{\frac{1}{2}}=0\,,
\label{03-bol01}
\end{equation}
can have two solutions. One of them is the point $K_a$ where
$G(\rho)$ vanishes. In the case, the trajectory of light always touch
the singular point, we will not consider this solution.
The other solution comes from
\begin{equation}
Q(\rho_0)=\frac{\rho_0^2G(\rho_0)}{F(\rho_0)}=\frac{L^2}{E^2}\,.
\end{equation}

The solutions of this equation exist in specific range of constant
$a$ and can be found by numerical calculation. To get the range,
we observe $Q(\rho)$ is asymptotically(large $\rho$) proportional to
$\rho^2$. With the behavior of asymptotic region,
the term $Q(\rho)$ must be zero at the point $K_a$ to cover all
region of positive values. Thus, the range of constant $a$ which
gives the region is
\begin{equation}
a<-1\ {\rm or}\ a>2 . \nonumber
\end{equation}

\begin{figure}
\centering
\includegraphics[bb=88 4 376 182,scale=0.6,keepaspectratio]{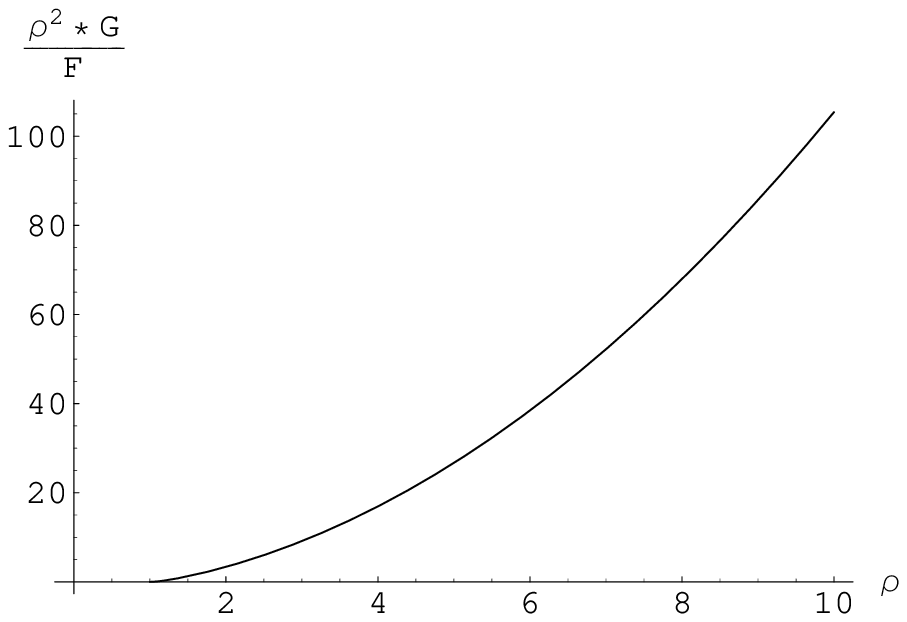}
\includegraphics[bb=61 4 391 36,scale=0.6,keepaspectratio]{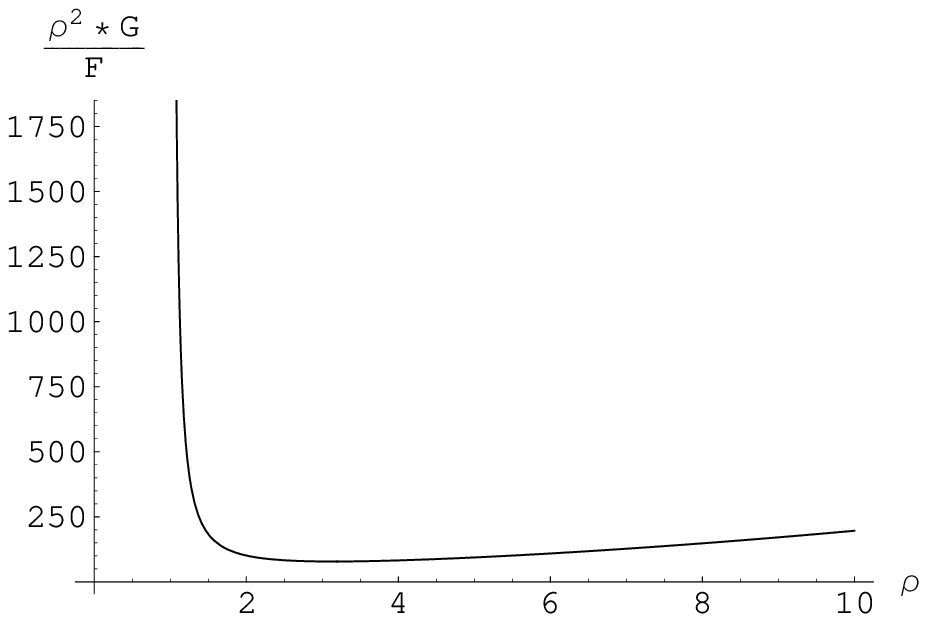}
\caption{The graph of $Q(\rho)$. The left graph is one of graphs
in range of $a<-1$ or $a>2$. The right graph is one of graphs in
range of $-1<a<2$.} \label{fig:curveap10}
\end{figure}

\begin{table}
\begin{center}
\begin{tabular}{|r|l|l|l|l|l|l|}  \hline
& a=-10 & a=-5 & a=-3 & a=4 & a=5 & a=10\\ \hline L=1 & 1.4472 &
1.2999 & 1.14769 & 1.14769 & 1.2365 & 1.42959\\ \hline L=2 &
2.15922 & 1.91702 & 1.61759 & 1.61759 & 1.80134 & 2.13164\\ \hline
L=3 & 3.01022 & 2.72839 & 2.36845 & 2.36845 & 2.59145 & 2.97837\\
\hline
\end{tabular}
\end{center}
\caption{The table of $\rho_0$ in the range $a<-1$ or $a>2$.}
\end{table}

The angle $\Delta\phi$ is expressed by
\begin{equation}
\phi(\rho_0)-\phi_{\infty}=\int^{\infty}_{\rho_0}\frac{L}{\rho^2G(\rho)^{\frac{1}{2}}}
\left(\frac{E^2}{F(\rho)}-\frac{L^2}{\rho^2G(\rho)}\right)^{-\frac{1}{2}}\,.
\label{03-bol01}
\end{equation}
This equation is redefined to make dimensionless as
\begin{equation}
\tilde{\rho}=K_a\rho,\ \ \ \tilde{L}=K_aL\,. \nonumber
\end{equation}
From now on, we drop tilde for simplicity.
The case of light is set to
$E=1\,$. The value of $\Delta\phi$ is relatived to
$\phi(\rho_0)-\phi_{\infty}$ by
\begin{equation}
\Delta\phi=2|\phi(\rho_0)-\phi_{\infty}|-\pi\,\label{32}.
\end{equation}
Using above table and Eq.\ (\ref{32}), the deflection angle $\Delta{\phi}$
is computed for some values of $a<-1$ or $a>2\,$ in table 2.

\begin{table}[h]
\begin{center}
\begin{tabular}{|r|l|l|l|l|l|l|}\hline
& a=-10 & a=-5 & a=-3 & a=4 & a=5 & a=10\\ \hline L=1 &          &
-1.12253 & -1.22649   & -1.22649  & -1.14726 & -1.10551\\ \hline
L=2 & -2.17775 & -0.24806 & -0.0552344 & -0.055234 &          &
-0.375315\\ \hline L=3 &          & 0.030895 & 0.295974   &
0.295974  & 0.124414 & -0.118769\\ \hline
\end{tabular}
\end{center}
\caption{The deflection angles in $a<-1$ or $a>2\,$.}
\end{table}

\begin{table}[h]
\begin{center}
\begin{tabular}{|c|c|c||c|c|c||c|c|c||c|c|c|}\hline
\multicolumn{3}{|c||}{a=-1}&\multicolumn{3}{|c||}{a=0}&\multicolumn{3}{|c||}{a=0.5}&\multicolumn{3}{|c|}{a=2}\\ \hline
L&$\rho_0$&$\Delta\phi$&L&$\rho_0$&$\Delta\phi$&L&$\rho_0$&$\Delta\phi$&L&$\rho_0$&$\Delta\phi$\\ \hline
4.12&1.41157&4.32406&9.8&5.15716&2.03338&12.3&7.13644&1.57997&4.12&1.41157&4.32406\\ \hline
7&4.79129&0.903441&12&7.8105&1.12041&16&11.3127&0.847878&7&4.79129&0.903441\\ \hline
\end{tabular}
\end{center}
\caption{The table for $-1\le a \le 2$ is computed.}
\end{table}

\subsection{The Timelike Geodesics}

For the timelike geodesic, $\mathcal{L}=-1$. Using Eq.\
(\ref{03-eula}) and Eq.\ (\ref{03-kvpmc01}) the equation becomes
\begin{equation}
g_{00}'\left(\frac{E}{g_{00}}\right)^2-g_{11}'\left(\dot{\rho}\right)^2
+ g_{33}'\left(\frac{L}{g_{33}}\right)^2
+g_{44}'\left(\frac{W}{g_{44}}\right)^2-2g_{11}\ddot{\rho} =0\,,
\label{03-tgc01}
\end{equation}
where the prime denotes the differentiation with respect to
$\rho\,$.

From Eq.\ (\ref{03-lagrangian}) we get
\begin{equation}
\dot{\rho}=\left(-\frac{1}{g_{11}}\left(1+\frac{E^2}{g_{00}}+\frac{L^2}{g_{33}}
+\frac{W^2}{g_{44}}\right)\right)^{\frac{1}{2}}\,.
\end{equation}
Inserting this result into the geodesic equation of $\rho$, we obtain
\begin{equation}
\ddot{\rho}=\frac{1}{2g_{11}}\left(\frac{g_{00}'}{g_{00}}\frac{E^2}{g_{00}}
+\frac{g_{00}'}{g_{11}}\left(1+\frac{E^2}{g_{00}}+\frac{L^2}{g_{33}}
+\frac{W^2}{g_{44}}\right)+\frac{g_{33}'L^2}{g_{33}^2}+\frac{g_{44}'L^2}{g_{44}^2}\right)\,.
\end{equation}
We get the relations between coordinate components as follows
\begin{eqnarray}
\frac{dt}{d\rho} &=& \frac{\frac{E}{g_{00}}}{(-\frac{1}{g_{11}}
(1+\frac{E^2}{g_{00}}+\frac{L^2}{g_{33}}+\frac{W^2}{g_{44}}))^{\frac{1}{2}}}\,,
\,\,\nonumber \\
\frac{d\phi}{d\rho}
&=&\frac{\frac{L}{g_{33}}}{(-\frac{1}{g_{11}}(1+\frac{E^2}{g_{00}}
+\frac{L^2}{g_{33}}+\frac{W^2}{g_{44}}))^{\frac{1}{2}}}\,, \,\,
\frac{dz}{d\rho}=\frac{\frac{W}{g_{44}}}{(-\frac{1}{g_{11}}(1+\frac{E^2}{g_{00}}
+\frac{L^2}{g_{33}}+\frac{W^2}{g_{44}}))^{\frac{1}{2}}}\,.
\nonumber \label{rdm1}
\end{eqnarray}

The timelike case allows stable circular orbits in some range of the
constant $a$. The effective potential for timelike geodesics is
\begin{eqnarray}
V_{eff}&=&\frac{1}{2}\frac{F(\rho)L^2}{\rho^2G(\rho)}+\frac{1}{2}F(\rho)
\nonumber \\ 
&=&V_{1}+V_{2}\,.
\end{eqnarray}

\begin{figure}[h] 
  \centering
  \includegraphics[bb=88 4 376 182, scale=0.6,keepaspectratio]{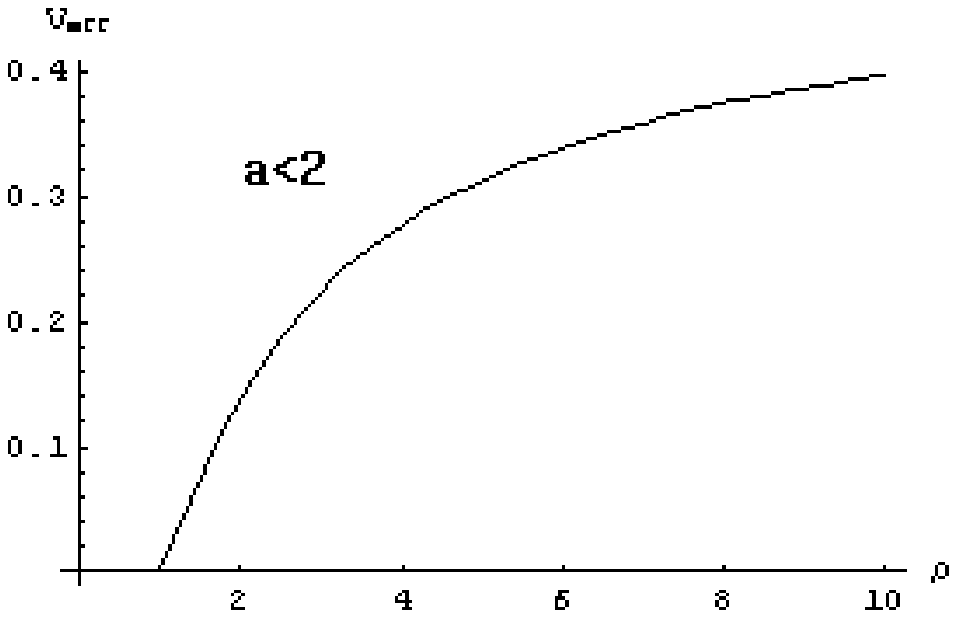}
\includegraphics[bb=88 4 376 182, scale=0.6,keepaspectratio]{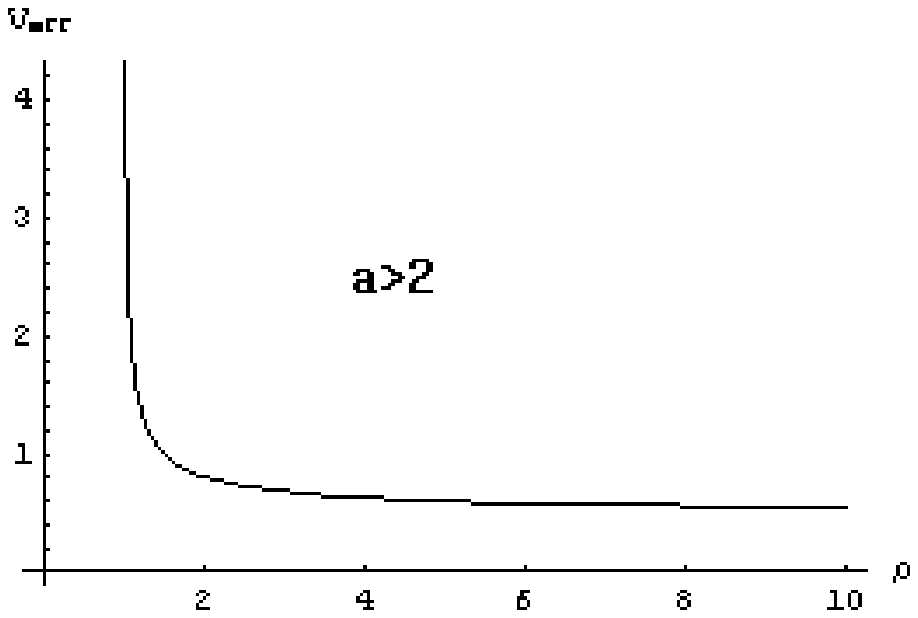}  
\caption{The left is a curve in $a<2\,$. Effective potential for L=0 in the potential
becomes zero at $\rho=K_a$ for $a<2\,$ while it becomes infinity at $\rho=K_a$ for $a>2\,$
. just fallen to
the singular point $K_a$. The right is in $a>2$. The potential wall exists in the point $\rho=K_a\,$.}
  \label{fig:timelikeradial1}
\end{figure}

First, we investigate radial motion of the effective potential
L=0 in Eq.\,(\ref{rdm1}) The behavior of the effective
potential is shown in Fig.\,\ref{fig:timelikeradial1}.

The potential is different from the null case due to the second
term for not $L=0\,$. To get the stable circular orbits, the asymptotic behaviors of
$V_{1}$ and $V_{2}$ must be analyzed. In asymptotic region,
$\rho\gg1$, each term becomes as follows
\begin{eqnarray}
V_{1}&\rightarrow&0 , \nonumber \\
V_{2}&\rightarrow&\frac{1}{2}\,.  \label{03-nt01}
\end{eqnarray}
As we can see in Eq.\,(\ref{03-nt01}), the asymptotic values are not affected
by the constant $a$. So the key conditions may be the behaviors of
$V_{eff}$ in the vicinity of $K_a$. At the point of $\rho=K_a$, a
convergent condition of $V_{2}$ is $a>0$. The calculation shows that
$V_{2}=0$ at $\rho=K_a$ in $a<2\,$.

\begin{figure}[h]
\centering
\includegraphics[bb=88 4 376 182,scale=0.6,keepaspectratio]{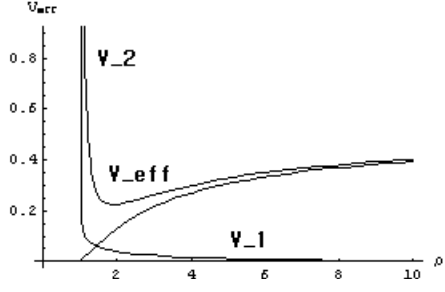}
\caption{Shape of the effective potential for the existence od the circular orbit.}
\label{fig:all3}
\end{figure}

For the existence of closed orbit, we require $V_1$ be divergent at $\rho=K_a$,
which gives the exponent $-2+s+\frac{1+a}{2-a}s$ being negative. this gives
$a<-1$ or $a>2$. We also want $V_2$ becomes zero at $\rho=K_a$ which gives $a<2\,$.
Thus, overlapping range for these conditions is $a<-1$.
In this region, the potential always has its minimum point, and the matter can be
moved following stable orbits. One of the circular orbit cases in the timelike potential
can be made from combination of $V_1$ and $V_2$ as shown in Fig. \ref{fig:all3}. Finding the condition of $a$ which exists the combination
at the same time is the key point in its analysis. Note that any angular momentum value is allowed.

\begin{figure}[h]
\centering
\includegraphics[bb=88 4 376 182,scale=0.6,keepaspectratio]{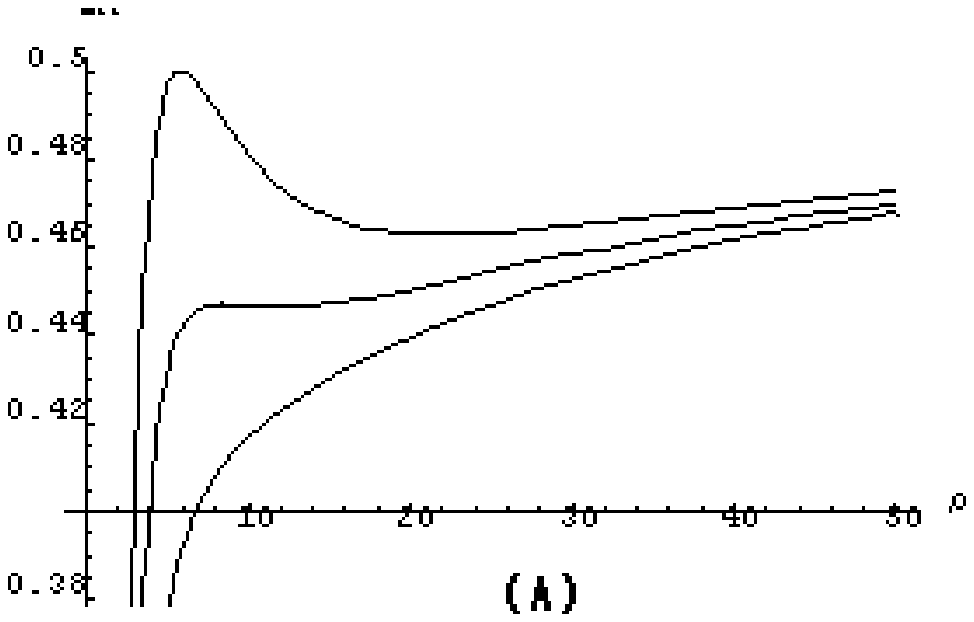}
\includegraphics[bb=88 4 376 182,scale=0.6,keepaspectratio]{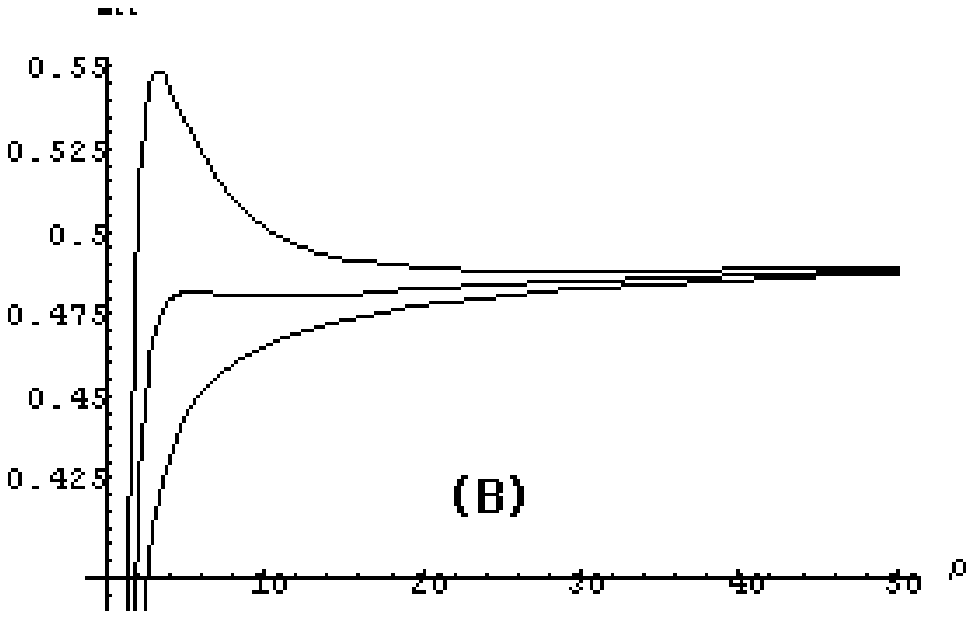}
\\
\includegraphics[bb=88 4 376 182,scale=0.6,keepaspectratio]{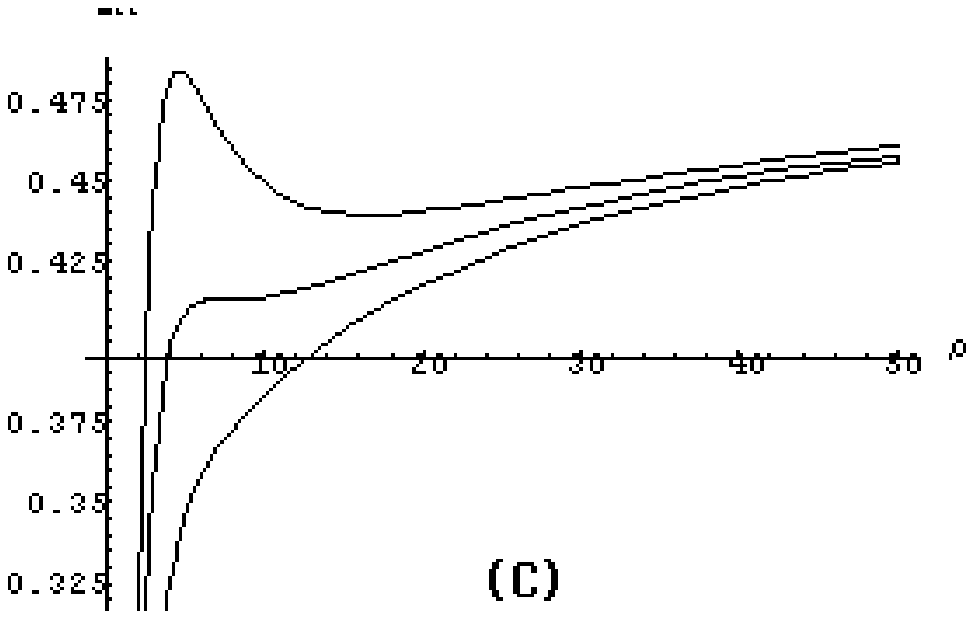}
\includegraphics[bb=88 4 376 182,scale=0.6,keepaspectratio]{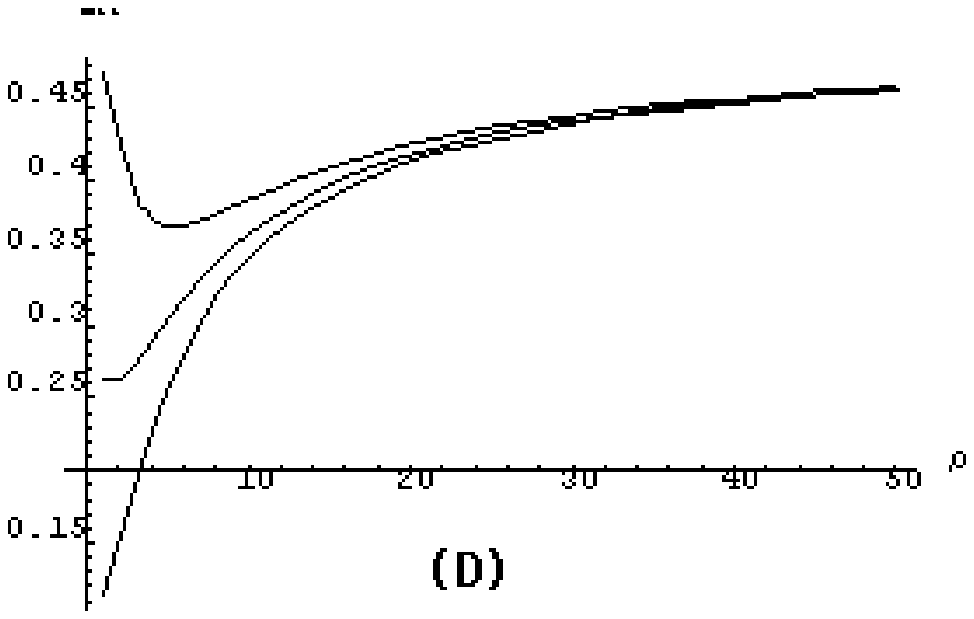}

\caption{(A)\,The case of the Schwarzschild
black string, $a=\frac{1}{2}\,$. In (A), upper curve is at $L=8$, and the
middle is at $L=7\,$, and the lower is at $L=6$(all figures are some
ordering.). The marginal orbit appears at middle curve. (B) is of $a=\frac{3}{2}$.
The angular momentums, $L\,$, are 4,
3, 2. (C) is for $a=0\,$. From above table 1, the radius of marginal orbit is
moved to left side and does not change its shape. (D)
is for $a=-1\,$. In this case, the shape of potential is different
from other case.} \label{fig:marginalorbit05total}
\end{figure}

In the range of $-1<a<2\,$, the orbit can exist only for
$r \geq 6M$ as in the Schwarzschild spacetime case if $a=\frac{1}{2}$. The orbit at $r = 6M$
is called the marginal stable circular orbit or the innermost
stable circular orbit. Some exaples are shwon in Fig. \ref{fig:marginalorbit05total}.

\begin{table}
\begin{center}
\begin{tabular}{|r|r|r|r|r|r|r|}\hline
$a$    &-1     &-0.5   &0      &0.5    &1      &1.5 \\    \hline
$L$    &2.9    &4.8    &6.4    &7      &5.2    &3   \\   \hline
$\rho$ &1.56673&4.59308&7.54371&8.38068&9.07818&5.48575 \\ \hline
\end{tabular}
\end{center}
\caption{The relations of $a$, $L$, $\rho$ are shown. For each
value of the constant $a\,$, the angular momentum $L$ and radius $\rho$ for marginal orbit are shown. In these analysis, objects which are not
$a=\frac{1}{2}$ have different geodesic properties, which
Schwarzschild black string has.}
\label{aaa12}
\end{table}

From the Fig.\ \ref{fig:marginalorbit05total}, we see that there are finite
potential barrier due to the angular momentum for the cases of
$a=0$, $a=\frac{1}{2}$ and $a=\frac{3}{2}$. On the other hand, the
potential barrier preventing a particle reaching deep inside appears appeared for the case of $a=-1$.

For some specific values of constant $a$, the
effect of angular momentum and marginal orbit is shown at Fig.\
\ref{fig:marginalorbit05total}, and this table\,\ref{aaa12} shows marginal
orbit's radius and values of angular momentum. As can be seen in the table \ref{aaa12}, the angular momentum which makes the marginal orbit becomes largest for $a=\frac{1}{2}$ while the radius of marginal orbits becomes the largest
approximately $a=1\,$.

\section{Summary and Discussions}

In this paper, we have studied the geodesic motions and the orbits of
both a massive particle and light ray. The geometry of the hypercylindrical
solution is dependent on single constant $a\,$, a ratio of tension and
mass density. This geometry becomes that of the
Schwarzschild black string for $a=\frac{1}{2}\,$, and the static
Kaluza-Klein bubble for $a=2$. There exist five conserved
quantities corresponding to translation symmetry of 
time, angle, 5th dimension coordinate, and two quantities which give equatorial plane $\theta=\frac{1}{2}\pi\,$. The quantities are E, L, and W related
to time, angle, and 5th coordinate. The geodesic equations for null
and timelike case is given by Euler-Lagrangian equations.
To get orbits, we obtained the effective potential from the
Lagrangian. We get effective mass from the equation. Effective mass is asymptotically flat, but it becomes zero at $\rho=K_a$ in $a<2\,$, and infinite in $a>2\,$. The light can move around a unstable
circular orbit in $-1\le a\le 2$. The radial range of the
unstable circular orbit is related to area of light capture. The capture cross section is formed in $-1\le a \le 2\,$, and the largest area case is $a=\frac{1}{2}\,$ Schwarzschild black string. The property of the metric at $\rho=K_a$ in our paper has already been studied by the authors in Ref.\ \cite{lkk}. The curvature of the metric was shown to be singular at the point  $\rho=K_a$ except for $a=1/2$ and $a=2$ in the range of $-1 <a <2$.
The geometry described by the hypercylindrical solution
affects null trajectories. The deflection angle is obtained to show
this effect. We get two range of constant a where null trajectories behave differently. The deflection angles are similar to black hole case in $-1 \le a \le 2\,$ and behave differently in $a<-1$ or $a>2\,$. There exist parameter range in which singularity is weakly naked one, $-1<a<2$, and strongly naked one, $a\leq-1$ or $a\geq2$, in our analysis. We calculate the timelike geodesic equations and the
range of the constant $a$ which gives stable circular orbits in $a<-1\,$. One
of the characteristics of the timelike case is that there exist a marginal
stable circular orbit in $-1<a<2$. The angular
momentum and radius of this marginal orbit is numerically
obtained, and the shapes of the effective potentials are similar to Schwarzschild black string in $-1<a<2\,$.

According to Virbhadra and Ellis, the singularity of the metric studied in the present work and in the Refs.\ \cite{chul, ckkl, ckkl1, hjko} corresponds to a weakly naked one. In other wards, Fig.\ \ref{fig:allplotm1} in our paper indicates that there exist photon spheres \cite{cvir} from $-1<a<2$. The observational properties of our spacetime is indistinguishable from the Schwarzschild black hole \cite{vire}. They may play the role of more efficient cosmic telescopes, if these singularities exist in nature \cite{virk}. The authors in Ref.\ \cite{virk} modeled the massive dark objects at galactic centers as these singularities including a Schwarzschild black hole.

With all these and in relation to the observation of the singularity in the future, it is worthwhile to investigate for the properties of the hypercylindrical spacetime. For instant, the scattering of waves is one of them.
Based on the information about the reflected wave, the reflecting rate and
scattering cross section can be calculated. This work is in progress.

\section*{Acknowledgments}

We would like to thank Gungwon Kang, Hyeong-Chan Kim, Chanyong
Park and Myungseok Yoon for valuable discussions and kind
comments. We would like to thank Jungjai Lee, Inyong Cho, and Chan-Gyung Park for
their kind comments at the Workshop on Numerical Relativity, June, 2008. This work was
supported by the Science Research Center
Program of the Korea Science and Engineering Foundation through
the Center for Quantum Spacetime (CQUeST) of Sogang University
with grant number R11 - 2005- 021. WL was supported by the Korea Research
Foundation Grant funded by the Korean
Government(MOEHRD)(KRF-2007-355-C00014).

\end{document}